# Direct positron emission imaging: ultra-fast timing enables reconstruction-free imaging

**Authors:** Ryosuke Ota[1]†, Sun Il Kwon[2]†, Eric Berg[2]†, Fumio Hashimoto[1], Kyohei Nakajima[3], Izumi Ogawa[3], Yoichi Tamagawa[3], Tomohide Omura[1], Tomoyuki Hasegawa[4], Simon R. Cherry[2]*

**Affiliations:**

[1]Central Research Laboratory, Hamamatsu Photonics K.K.; Hamamatsu, Japan

[2]Department of Biomedical Engineering, University of California; Davis, USA

[3]Faculty of Engineering, University of Fukui; Fukui, Japan

[4]School of Allied Health Science, Kitasato University; Kitasato, Japan

†These authors contributed equally to this work.

*Corresponding author. Email: srcherry@ucdavis.edu

**Abstract:** Positron emission tomography, like many other tomographic imaging modalities, relies on an image reconstruction step to produce cross-sectional images from projection data. Detection and localization of the back-to-back annihilation photons produced by positron-electron annihilation defines the trajectories of these photons, which when combined with tomographic reconstruction algorithms, permits recovery of the distribution of positron-emitting radionuclides. Here we produce cross-sectional images directly from the detected coincident annihilation photons, without using a reconstruction algorithm. Ultra-fast radiation detectors with a resolving time averaging 32 picoseconds measured the difference in arrival time of pairs of annihilation photons, localizing the annihilation site to 4.8 mm. This is sufficient to directly generate an image without reconstruction and without the geometric and sampling constraints that normally present for tomographic imaging systems.

**Introduction:**

Three-dimensional biomedical imaging techniques including x-ray computed tomography (CT) single photon emission computed tomography (SPECT) and positron emission tomography (PET) measure one-dimensional or two-dimensional projections from the object of interest that subsequently are reconstructed into cross-sectional images or 3-D image volumes via analytic computed tomography algorithms based on the Radon transform (1), or using iterative algorithms, typically based on the expectation-maximization algorithm (2). In magnetic resonance imaging (MRI), data are natively acquired in the frequency domain and are subsequently reconstructed into images in the spatial domain through the Fourier transform (3). In all these imaging modalities, a measured data point does not have a 1:1 correspondence with a point in image space, and the spatial distribution of the signal must be inferred by a reconstruction step, which amplifies noise levels. Accurate tomographic image reconstruction also depends on adequate angular (PET, SPECT and CT) or frequency (MRI) sampling of the data.

Uniquely among these imaging modalities, PET can localize the signal source beyond the entire line of response by exploiting the time difference in detection of the two back-to-back annihilation photons produced following the emission of a positron (Fig. 1). This is the basis for time-of-flight PET (Fig. 1A), and the best systems currently available have a timing resolution of ~210 ps (4) resulting in a spatial localization along the line of response of 3.15 cm. As shown in Fig. 1C, this constrains the possible location of a detected event, but does not define the source location, and therefore image reconstruction with all its concomitant limitations is still required. The propagation of noise from the reconstruction algorithm, and the predicted improvements in image signal-to-noise ratio (SNR) as a function of time-of-flight resolution, have been previously studied as summarized by (5).

Once the timing resolution becomes sufficiently good enough to directly localize the source, we enter a new regime, in which an image can be directly obtained without any reconstruction step. We call this direct positron emission imaging (dPEI) (Fig. 1D). This approach is somewhat analogous to ultrasound, which also uses time-of-flight differences, at the speed of sound in tissue, to localize the depth of ultrasound-reflections. However, in dPEI, the time-of-flight differences are governed by the speed of light rather than the speed of sound, resulting in time differences of tens of ps rather than μs. In this work we developed very fast radiation detectors, with an average coincidence timing resolution of 32 ps, and demonstrate, for the first time, the generation of cross-sectional images of the distribution of a positron-labeled radiotracer while completely eliminating the noise-amplifying image reconstruction algorithm.

Here, we describe the enabling technological and methodological innovations, measure the timing performance and the relationship between the source localization and the measured time-of-flight difference, and show cross-sectional images of three different test objects, produced directly from a single angular view without tomographic image reconstruction.

**Approach:**
Fast time-of-flight radiation detectors used for imaging positron-emitting radionuclides normally consist of a bright, high-density scintillator coupled to a silicon photomultiplier (SiPM) that electronically converts and amplifies the scintillation light through the generation of electron-hole pairs in the semiconducting material (6). However, the rate at which photons are produced by the scintillation process is relatively slow, due to the time needed to form excited states and for recombination to occur (7). Furthermore, silicon photomultipliers have a single photon time resolution (SPTR) that is typically on the order of 100-300 ps (8,9). Therefore, to achieve timing resolutions of 40 ps or better needed for dPEI likely requires a different strategy. Various approaches have recently been discussed as part of an international challenge launched to focus efforts on ultimately achieving 10 ps timing resolution (10). In this work we combined three innovations to make dPEI possible, namely the use of Cerenkov luminescence as the mechanism to achieve a fast timing signal, the integration of a Cerenkov radiator directly within the photosensor to optimize light transport and photodetection timing properties, and the application of convolutional neural network (CNN) as a standalone algorithm to predict the timing information from the measured detector waveforms.

Cerenkov radiation is produced when a charged particle travels faster than the phase velocity of light in a medium, is emitted promptly, and therefore presents as an attractive mechanism to be exploited for ultra-fast timing applications (11,12). The detection of 511 keV photons emitted following positron-electron annihilation in materials with high refractive index and high atomic number create sufficiently energetic electrons to produce a small number of Cerenkov photons (13–15). Next, this prompt optical signal needs to be converted to an electronic signal and amplified, and for this we developed a photosensor based on the structure of microchannel plates (MCP). MCP photomultiplier tubes (MCP-PMTs) are known for their outstanding SPTR, with values that can approach 20 ps (16). For this work, we developed an MCP-PMT in which the Cerenkov radiator ($PbF_2$ glass) was integrated with the photocathode inside the MCP-PMT (Fig. 2A), thus removing all optical boundaries that had hampered detection of the Cerenkov photons and reduced detection (17). These devices are henceforth referred to as Cerenkov radiator integrated MCP-PMTs or CRI-MCP-PMTs. We also further refined the design by removing lead-based compounds from the MCP structure to reduce the probability of direct interactions of the 511 keV photons which leads to side peaks in the timing spectrum and ultimately results in ambiguous localization for a small percentage of detected events (18). Lastly, we used a convolutional neural network (CNN) to determine the time of flight difference for detected events, extending previously developed methods (19). By placing radioactive point sources at different locations between a detector pair, large numbers (>$10^6$) of training events with known ground truth time-of-flight differences can readily be collected for training, allowing the CNN to learn the complex waveforms and output the time-of-flight difference. The CNN is a 9-layer network where each layer contains convolution, batch normalization, and a rectified linear activation function. The whole CNN was trained using stochastic gradient descent.

**Results:**
All data was acquired using two CRI-MCP-PMTs, placed in coincidence, with lead or tungsten collimation of the 11 mm diameter active area of the photocathode based on the desired image resolution. After optimizing the bias voltages supplied to the different stages of the CRI-MCP-PMTs the estimated SPTR was 22 ps, and the gain was ~$1.8\times10^6$ when a bias voltage of -3.0 kV was supplied. The CRI-MCP-PMT has the same quantum efficiency (QE) curve as the R3809U-50 MCP-PMT (Hamamatsu Photonics K.K., Japan) with a QE above 20% for the wavelength range 200-420 nm. Fig. 2B shows a histogram of the time of flight differences for a point source of $^{22}$Na located at the center of a detector pair. The measured coincidence timing resolution based on constant fraction timing of the digitized waveforms was 32.9 ps, improving to 26.4 ps by using the CNN. Fig. 2C shows the time-of-flight histograms for 5 source locations, spaced by 25 mm, demonstrating the linear relationship between the measured time-of-flight different and source location (also see Fig. S5), as well as the relatively uniform timing resolution achieved across a 10 cm range (26.4 – 35.7 ps). The corresponding spatial resolution was 3.96 mm at the center, and was better than 5.36 mm across the entire range.

Imaging studies were performed in three different test objects filled with an aqueous solution of the radiotracer $^{18}$F-fluorodeoxyglucose ($^{18}$F-FDG) ($T_{1/2}$=110 mins.). To capture images, the detector pair was translated linearly to cover the width of the object, building up the image one row of pixels at a time (Fig. 3). The acquisition duration at each position was adjusted for radioactive decay to provide roughly equal counting statistics for each measurement. Images were generated directly from the measured data without any reconstruction, using the position of the detector pair to determine the *x*-coordinate and the time-difference of the two detected events to determine the *y*-coordinate. No rotation of the object or detectors was required to form a cross-sectional image. Fig. 3C shows the first dPEI image produced.

Figure 4 shows the images for each of the test objects using the CNN to determine the timing difference on an event-by-event basis, and correcting for radioactive decay, acquisition time, and attenuation of the 511 keV photons by the object based on analytic calculations. Each of the test objects highlights a different imaging attribute or task. The first object (Fig. 4A) is commonly used as an image quality test in preclinical PET scanners and consists of a uniform background with two voids where there is no activity, one filled with air, and one filled with water. Both voids, which are 8 mm in diameter, are clearly visualized. The second object is a resolution test pattern with radioactive rods of different diameters and a spacing equal to twice their diameters (Fig. 4B). The image demonstrated that the 3 mm rods can be resolved, indicating that a spatial resolution on the order of 4-5 mm is recovered in the dPEI images in line with expectations based on the average measured timing response of 32 ps. The third object is much larger (18.4 cm in diameter) and represents the distribution of FDG in a slice of the human brain (Fig. 4C). The detailed structure in this object is faithfully captured in the dPEI image with a spatial resolution of ~4.8 mm and demonstrates that the method could be scaled for human imaging.

**Discussion:**

These images represent the first examples of the direct localization and imaging of a positron-emitting radionuclide, using data from a single angular view and without any tomographic image reconstruction algorithm, to generate a cross-sectional image. This two-detector system with its average timing resolution of 32 ps was able to produce images at a spatial resolution of 4.8 mm in the timing direction. This spatial resolution is already similar to that achieved in images from TOF-PET scanners used for diagnostic purposes. The spatial resolution in the *x*-direction is roughly one-half the collimated detector width. 3D volumetric images could in principle be acquired by also translating the detector pair in the *z* direction.

A number of limitations must be addressed in order to develop more practical dPEI systems. The acquisition times in these first imaging experiments were long (2-34 minutes per measurement position, 4-24 hours for the whole image) and the amount of radioactivity used high (up to ~1000 MBq). Fortunately, there are obvious avenues to increase signal collection efficiency. These include using a higher atomic number radiator (such as the scintillator bismuth germanate, which produces both Cerenkov and scintillation light) in place of the lead glass radiator used in these MCP-PMTs, increasing the thickness of the radiator to 4-5 mm, developing multi-channel detectors that can be tiled together, and then using multiple detectors arranged in linear arrays or panels to increase geometric coverage and collection efficiency.

These three changes alone (see supplementary materials) could easily increase the detection sensitivity by >$10^3$, reducing acquisition times or radiation doses accordingly. Multi-detector configurations that cover the imaging volume of interest would also remove the need for detector translation and allow dynamic imaging of radiotracer distributions.

dPEI as a biomedical imaging modality offers a number of interesting opportunities, by freeing the design of the imaging system from the typical constraints associated with the sampling necessary for tomographic reconstruction. For example, dPEI systems need only cover the region of an object that is of interest, as truncation artifacts, which are prevalent in tomography, are no longer an issue. Also, in the medical setting, systems need not consist of complete detector rings that enclose a subject, but can allow a more open geometry for improved access to, and acceptance by, patients, while still providing fully 3D images. In addition, because each individual event carries the information needed to completely localize it in 3D space, the image signal-to-noise ratio is maximized for a given number of detected events. This is exemplified by Fig. 4, where each image is comprised of just 4,000 – 20,000 events, yet this is sufficient to clearly visualize the object. Once larger scale systems, with more efficient detectors are developed, it should be possible for dPEI to acquire high SNR images at low radiation doses and with short acquisition times. Furthermore, dPEI opens up the possibility of real time imaging, as no reconstruction step is involved and images can be viewed as they are in the process of being acquired.

**Acknowledgments:** We thank Dr. Hiroyuki Ohba, Dr. Shingo Nishiyama and Dr. Masakatsu Kanazawa for their kind technical support at Hamamatsu Photonics, and George Burkett and Steven Lucero at the University of California Davis for fabricating the spatial resolution phantom and 3-D printed holders used in this study.

**Funding:**

National Institutes of Health grant R35 CA197608 (SRC, SIK, EB)

National Institutes of Health grant R03 EB027268 (EB, SRC)

**Author contributions:**

Conceptualization: SRC, RO, SIK, EB, FH, TO, TH

Methodology: SRC, RO, SIK, EB, FH, KN, IO, YT, TO, TH

Resources: RO, TO

Investigation: SIK, RO, EB, FH

Visualization: SIK, RO, EB

Funding acquisition: SRC

Supervision: SRC, TO, TH

Writing – original draft: SRC, SIK, RO, EB

Writing – review & editing: SRC, SIK, RO, EB, FH, KN, IO, YT, TO, TH

**Competing interests:** Authors declare they have no competing interests.

**Data and materials availability:** The raw detector waveform data from this study, as well as the code to read that data and apply the trained CNN are available by request to the authors and a completed data transfer agreement.

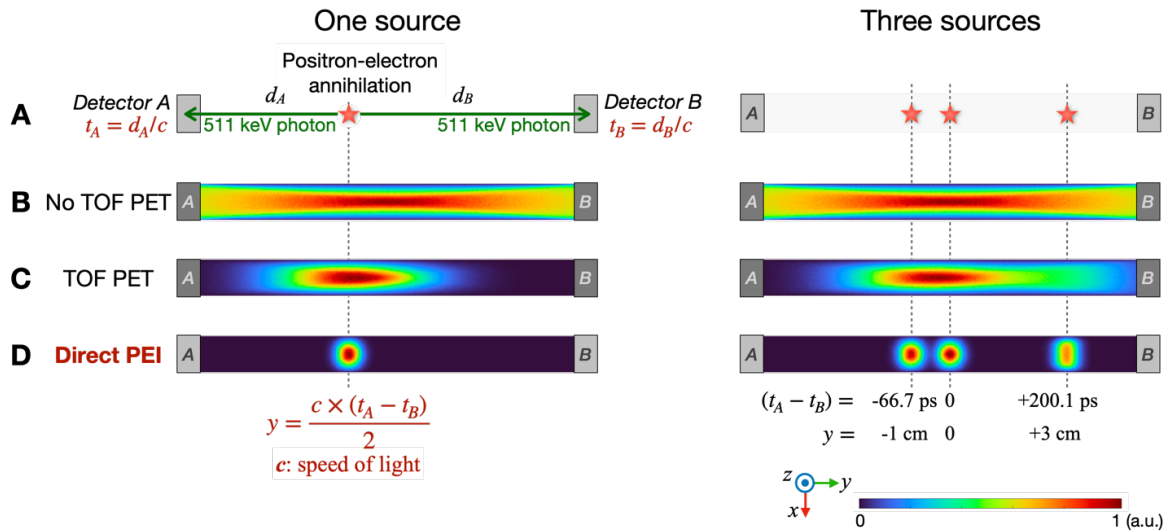

**Fig. 1. Basis for direct imaging of positron-emitting radiotracers using ultra-fast timing.** (A) Detection of back-to-back annihilation photons by a pair of radiation detectors and the source location-dependent arrival times of the two photons. (B) Simulations showing the probability distribution of source locations for a detected event (9.6 × 9.6 mm² detector area, detector separation = 96 mm) in conventional PET imaging where no timing information is available. When multiple sources are present (right hand side), it is not possible to discriminate them without additional measurements from different detector orientations. (C) partial localization when time-of-flight (TOF) information is added (example shown is 210 ps timing resolution, corresponding to 3.15 cm spatially); (D) Direct PEI, the new modality, in which the timing resolution of 32 ps allows the event to be localized within 4.8 mm, providing the basis for direct image generation without image reconstruction. The three sources can now be clearly resolved from a single measurement. The 2-D sensitivity maps presented are computed by simulation using the stated geometry and are calculated by integrating the 3-D distributions across the detector width perpendicular to the figure.

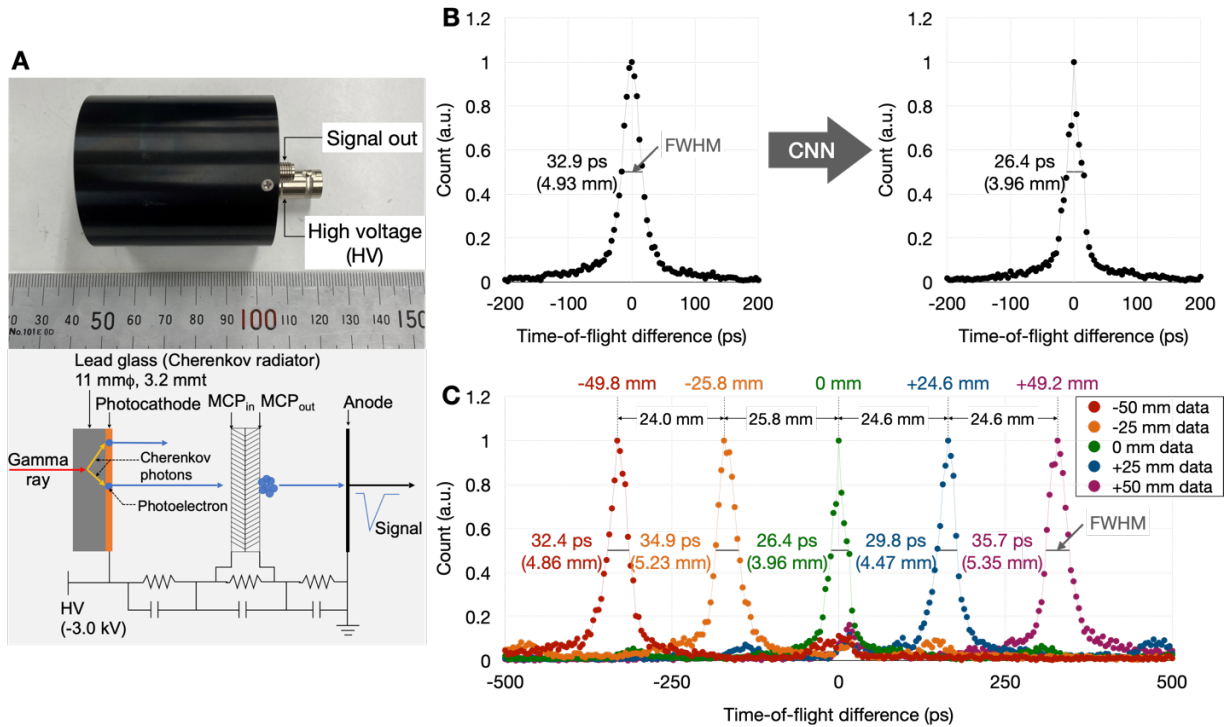

**Fig. 2. Timing resolution of 32 ps measured with MCP-PMT radiation detectors.** (A) Photograph and schematic of the microchannel plate photomultiplier tube (MCP-PMT) with an integrated lead-glass Cerenkov radiator as its entrance window, allowing Cerenkov radiation to reach the photocathode and liberate electrons without any intervening optical barriers. The MCP structure also is modified to remove lead compounds to reduce direct interactions of 511 keV annihilation photons in the structure itself; (B) Histogram of time-of-flight differences measured from two MCP-PMTs with integrated Cerenkov radiators in coincidence for a centrally-located $^{22}$Na point source. The MCP-PMT signals are digitized at 50 giga-samples-per-second, and the timing pick off determined by applying constant fraction discrimination in software. Using a convolutional neural network to estimate the time-of-flight difference of the waveforms further improves the timing resolution from 32.9 ps to 26.4 ps; (C) Histograms of time of flight differences as the point source is moved between the two detectors in steps of 2.5 cm, showing the linear relationship with source location. The calculated source location based on the timing difference is shown and is within 0.8 mm of the actual location. The timing resolution varied between 26.4 and 35.7 ps across the 10 cm range studied. The average timing resolution across the range was 32 ps. The timing resolution was measured as the full width at half maximum (FWHM) of the timing distributions.

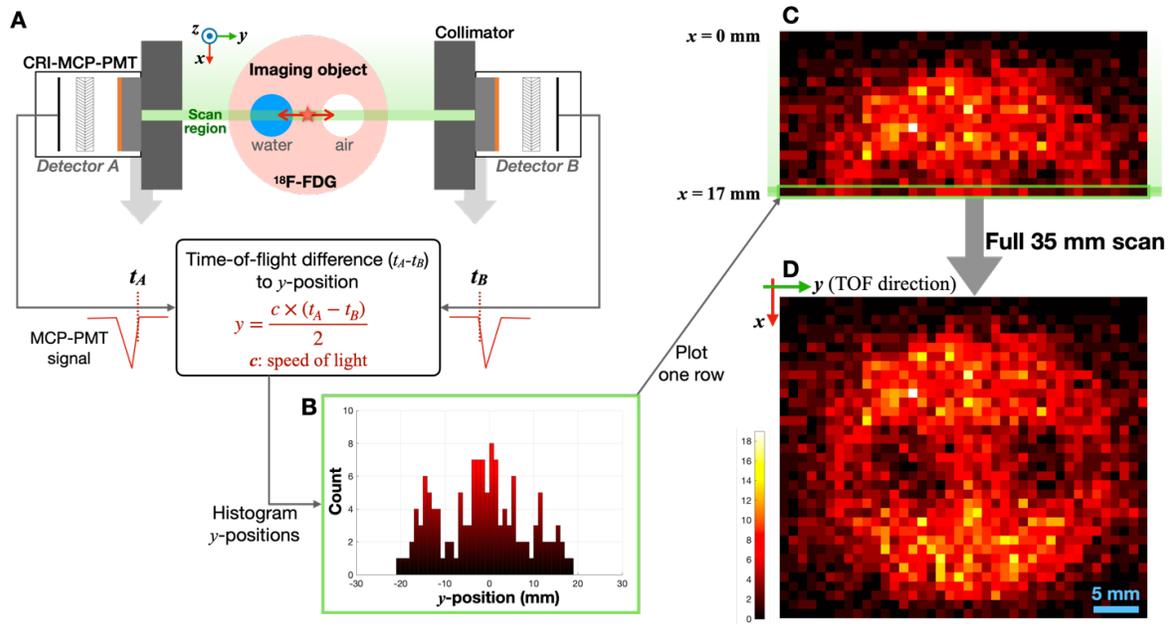

**Fig. 3. Acquiring a cross-sectional image using a pair of CRI-MCP-PMT detectors.** (A) The x-direction of the image is encoded by the position of the collimated detector pair, and the y-direction is encoded by the timing information as shown in Fig. 2D. The test object is filled with the radiotracer $^{18}$F-fluorodeoxyglucose ($^{18}$F-FDG), except for two voids (one air and one water); (B) Data is acquired for each x-position of the detector pair and the timing information used to determine the distribution of activity along the line between the two detectors; (C) The image is built up line by line as the detectors are translated. The image resolution in the x direction is governed by the opening of the collimator, and in the y-direction by the timing resolution of the detector pair; (D) The final raw image.

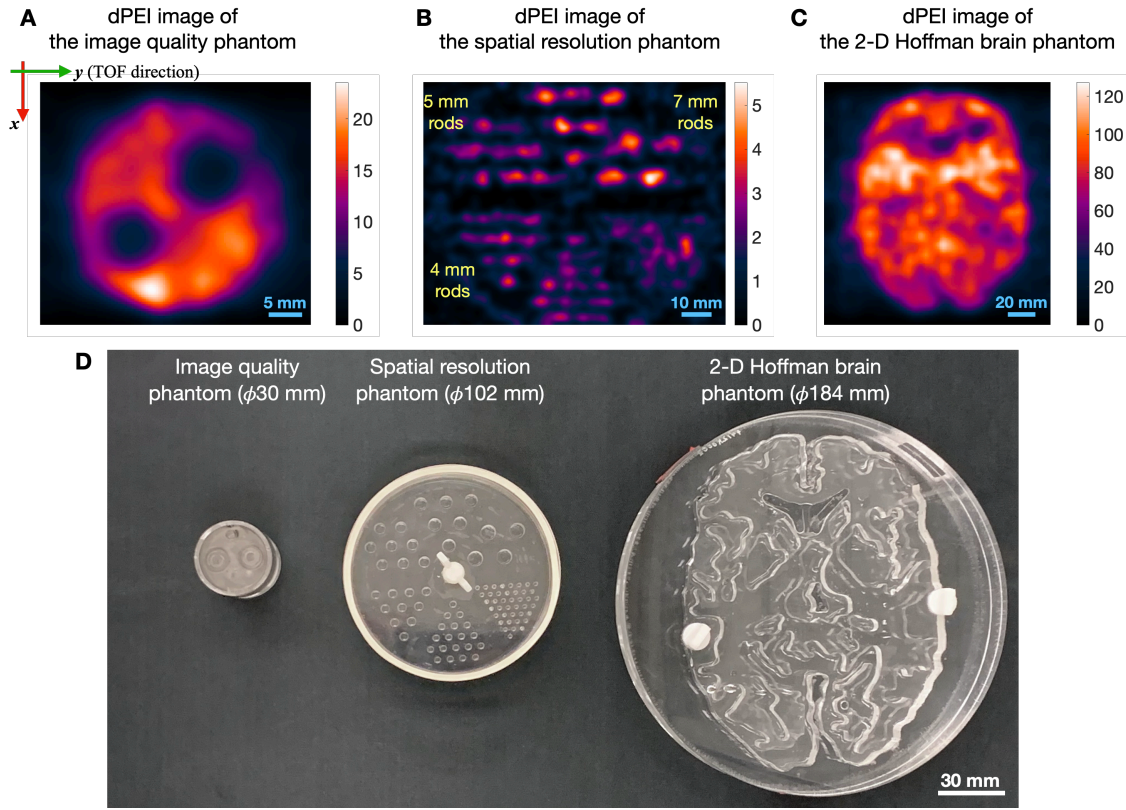

**Fig. 4. Cross-sectional images for different scale objects directly measured from a single angular view and without image reconstruction.** Images were acquired with the dPEI set up illustrated in Fig. 3. Corrections for radioactive decay, acquisition time, geometric sensitivity and photon attenuation have been applied to the raw image data; (A) Image quality test object with 8 mm diameter air and water-filled voids; (B) Structured object with an array of sources of different sizes and spacing to illustrate spatial resolution. In both (A) and (B), objects were imaged with a detector spacing of 20 cm, a collimator opening of 2 mm by 10 mm (the latter defining the imaging slice thickness) and the detectors were translated by 1 mm to ensure sufficient sampling in the x-direction. The y-direction was sampled in 6.67 ps increments, also corresponding to 1 mm. (C) The detectors were moved further apart (30 cm) and the collimation opened up to 8 mm by 10 mm to scale up to an object size with dimensions relevant to human imaging. dPEI images of the 2-D Hoffman brain phantom representing the distribution of $^{18}$F-FDG in a slice of the human brain acquired with 4 mm step size. (D) Corresponding photograph of the test objects showing dimensions. For details see supplementary materials.

# Supplementary Materials

**Materials and Methods**

Simulating the probability distribution of source locations for a detected event

Monte Carlo simulation studies were conducted to illustrate the differences in the probability distribution for source locations in coincidence event detection comparing positron emission tomography (PET), PET with time-of-flight capability (TOF-PET), and direct positron emission imaging (dPEI). GATE version 9.0 software (*20, 21*), which is based on the GEANT 4 simulation platform, was used. In the simulation studies, two radiation detectors were arranged face-to-face with a separation of 96 mm (Fig. 1A). The active area of each detector was $9.6 \times 9.6$ mm$^2$. The volume between the two detectors ($9.6 \times 9.6 \times 96$ mm$^3$) was uniformly filled with 1 MBq of activity, with the source being defined as an ideal back-to-back 511-keV mono-energetic emitter in air. A total of approximately 31 million coincidence events were acquired, with each event containing information about the source location and the interaction position on each detector surface.

The difference in arrival time of two photons in each coincidence event was calculated using the recorded source position and the two detected positions. For display purposes, the 2-D probability distribution of the estimated source locations on the *xy*-plane was calculated from the 3-D distribution by integrating the probability maps along the *z*-direction. For PET alone (Fig. 1B), where no timing information is available, the distribution is governed by coincidence detection geometry alone. For TOF-PET (Fig. 1C), the 2-D probability distribution was calculated by applying Gaussian weighting in the time domain with a timing resolution of 210 ps full-width at half-maximum (FWHM), which models the best-reported timing resolution for a commercial TOF-PET system (*4*). In the case of dPEI imaging, Gaussian weighting with a timing resolution of 32 ps FWHM, which models the timing performance measured for the dPEI system used in this study, was applied in the time domain. For the right-hand column of Fig. 1 where three adjacent sources were modeled, we applied: i) no time domain weighting for standard PET imaging, ii) three Gaussian weightings centered on the source location each with the same timing resolution of 210 ps FWHM for TOF PET imaging, and iii) three Gaussian weightings with 32 ps FWHM for dPEI imaging.

Cerenkov radiator integrated MCP-PMT (CRI-MCP-PMT) photodetector

A pair of microchannel plate photomultiplier tubes (MCP-PMTs), in which a lead glass Cherenkov radiator was integrated as the window face plate of the MCP-PMTs (referred to as

CRI-MCP-PMTs), were designed, fabricated and used to detect 511 keV photons in coincidence with ultra-high timing resolution. The dimensions of each MCP-PMT are 45 mm in diameter and 52.5 mm length, with an active detection area 11 mm in diameter (Fig. 2A). The lead glass window face plate is 3.2 mm thick. The MCP incorporated in the PMT structure consisted of borosilicate glass fabricated through an atomic layer deposition (ALD) technique that replaced conventional lead-based compounds to reduce the probability of 511 keV photons directly interacting within the MCP structure, instead of the Cerenkov radiator window faceplate. Such directly interacting events would otherwise degrade timing performance (*22*). A multialkali photocathode was deposited on the lead glass Cerenkov radiator via a 2-3 nm thick $Al_2O_3$ intermediate layer to protect against chemical reactions occurring between the lead glass and the photocathode when using the ALD technique.

To optimize the intrinsic timing performance of the MCP-PMTs, a custom voltage divider circuit was employed to tune the photoelectron gain across the MCP-PMT. Resistors of 7.5 MΩ, 18 MΩ, and 15 MΩ were selected between the photocathode and $MCP_{in}$, $MCP_{in}$ and $MCP_{out}$, and $MCP_{out}$ and the anode, respectively (Fig. 2A). A voltage of –3.0 kV was supplied to the photocathode of each MCP-PMT. The entrance surface of the MCP-PMT was covered by black tape to suppress internal optical reflections in the radiator to further optimize the detector's timing performance.

Methods, materials and data acquisition at Hamamatsu Photonics K.K.

Initial experiments were performed independently at both the Central Research Laboratory, Hamamatsu Photonics, Japan, and the University of California, Davis (UC Davis) to ensure reproducibility of the results. Experimental conditions for the CRI-MCP-PMT detectors (detector modules used, divider circuit, supplied bias voltage, black tape) were identical for both experimental sites, while there were minor differences in experimental setups used due to equipment and materials available. An overview of both experimental setups is shown in Fig. S1. A 50 mm thick lead collimator with 2 mm opening width was used for the experiments at Hamamatsu Photonics, K.K., Japan. The NEMA NU4 image quality phantom (Fig. S2A and Fig. 4D), a standardized object used for testing PET imaging systems (*23*), was fixed on a linear translation stage (SGSP20-35, Sigma Koki, Japan) and motorized by a stage controller (SHOT-102, Sigma Koki, Japan).

The NEMA-NU4 phantom was filled with 9 mL of $^{18}$F-NaF in aqueous solution and with an initial activity concentration in the background region of 227 MBq/ml (2.04 GBq total activity), as measured by a well counter (IGC-7, Aloka, Japan). One of the 8-mm inner diameter voids was filled with water, the other was filled with air. The phantom was stepped laterally across its width in 0.5 mm increments. 70 timing spectra covering 3.5 cm were collected in total.

Signals from the MCP-PMTs were directly fed into an oscilloscope (DSOS404A, Keysight, USA) and digitized at 20 GS/s with a bandwidth of 4.2 GHz. Waveforms were fully digitized on an event-by-event basis, and transferred to a personal computer for analysis. The trigger time difference ($t_A - t_B$) between the two digitized signals for each event was calculated and used to estimate the location of positron annihilation. The threshold level for the triggering each signal was set to 4% of the pulse height of the signal. The total duration of data acquisition was ~5.5 hours and the acquisition duration at each location was gradually increased from 2 mins to 15.6 mins to account for $^{18}$F decay (half-life = 110.9 minutes). The timing spectrum obtained for one specific measurement location is shown in Fig. 3B and the final raw image data produced from this scan is shown in Fig. 3D.

Methods and materials at UC Davis

The same two CRI-MCP-PMTs that were used in the study at Hamamatsu Photonics K.K. were transferred to UC Davis for comparative dPEI imaging studies of the same NU4 image quality phantom, along with several additional evaluations of their timing performance and extended imaging experiments on different test objects. The two CRI-MCP-PMT detectors were mounted on top of an optical breadboard (Thorlabs Inc., USA) with their line-of-response parallel to the *y*-axis. The custom holders for the two detectors were 3-D printed for precise alignment. In front of each detector, a custom adjustable tungsten collimator was fastened in front of each detector face, consisting of four 3.81 cm thick tungsten alloy cubes (Midwest Tungsten Service, USA). The four cubes were positioned such that their relative offsets determined the collimator height (slice thickness) along the *z*-axis, and width along the *x*-axis (Fig. S1B). The distance between the two detectors and the opening area of each collimator were varied depending on the object being imaged. A bias voltage of −3 kV was applied to each detector. The output signals output from the CRI-MCP-PMTs were digitized with an oscilloscope (DPO71254C, Tektronix, USA) at a sampling rate of 50 GS/s and with a bandwidth of 12.5 GHz. Coincidence events were determined by a sequential logic triggering function in the oscilloscope using cables with different lengths.

Experimental datasets

Four sets of experimental data were acquired at UC Davis.

*1) Point source experiments:* The purpose of this measurement was to characterize i) the relationship between measured time-of-flight differences and source location and ii) the coincidence timing resolution corresponding to the spatial resolution of the detector pair along the *y*-direction (see Fig. 3A for coordinate system). Coincidence events were acquired from a $^{22}$Na radioactive point source placed at 5 different source locations spaced by 25 mm along the detector line-of-response (*y*-axis). The distance between the two detectors was 300 mm, and the opening area of each collimator was 8 mm (*x*-direction) and 10-mm (*z*-direction). The point source activity was 4.1 MBq, and the acquisition duration at each point source position was ~3 hours to ensure collection of sufficient counting statistics.

*2) dPEI scan of the NEMA NU-4 image quality phantom:* A 1-D motorized translation stage (Velmex Inc., USA) was mounted on top of the optical breadboard between the two stationary detectors to scan test objects relative to the detector pair along the *x*-direction. The NEMA NU-4 image quality phantom was prepared (Figs. 4D and S2A) similarly as described previously. The 30-mm inner diameter uniform background region was filled with 847 MBq of the radiotracer $^{18}$F-fluorodeoxyglucose (FDG) in aqueous solution. The phantom was scanned using the custom 3-D printed holder shown in Fig. S1B to ensure accurate location and alignment. The distance between two detectors was set to 200 mm, and the opening area of each collimator was set to 2 mm (*x*-direction) and 10-mm (*z*-direction). The phantom was imaged by stepping along the *x*-direction 35 times with a 1-mm step size. The duration of data acquisition at each position was adjusted to account for $^{18}$F decay: the acquisition duration at the first *x*-position was 3.56 mins, and increased up to 15.1 mins for the last position. The total acquisition time to acquire the entire dPEI image was 4 hours.

*3) dPEI scan of a spatial resolution phantom*: A spatial resolution test phantom (also known as a Derenzo phantom) was fabricated from acrylic. The 102-mm outer-diameter phantom consisted of 6 sectors, each sector was composed of multiple 9.53-mm tall rods with triangular equilateral spacing equal to twice the rod diameter. The rod diameters were 2, 3, 4, 5, 6, and 7 mm, respectively (Fig. 4D and S2B). Each rod was filled with $^{18}$F-FDG using a syringe, and the total activity was 1,025 MBq. The distance between two detectors was 200 mm, and the opening area of each collimator was 2 mm (*x*-direction) by 10-mm (*z*-direction). The phantom was imaged by stepping along the *x*-direction 82 times with a 1-mm step size. Since the phantom size was larger than the image quality phantom, the acquisition time at the first *x*-position was reduced to 1.85 mins and the total scan time to cover the entire phantom area was 8 hours.

*4) dPEI scan of the 2-D Hoffman brain phantom:* A 2-D Hoffman brain phantom (Data Spectrum, USA) which represents the distribution of the radiotracer $^{18}$F-FDG in the human brain was imaged (Fig. 4D). The phantom has a diameter of 184 mm and a maximum water-fillable thickness of 13 mm. Regional contrast differences (gray matter:white matter:ventricle ratio = 4:1:0) are created through partial volume effects according to the slice thickness. To scan this large brain phantom, the distance between the two detectors was increased to 300 mm, and the collimator opening width was set to 8 mm (*x*-direction) and 10-mm (*z*-direction). The phantom was filled with ~850 MBq of $^{18}$F-FDG and was imaged by stepping along the *x*-direction 44 times with a 4-mm step size. The first acquisition time was 3.27 mins, and all other acquisition times were increased to adjust for $^{18}$F decay so that similar counting statistics were achieved in each measurement. For the last measurement position the acquisition time was 28.6 mins. The total scan time was 6 hours. We repeated this scan eight times by re-filling the phantom with radioactivity and carefully re-positioning it inside the imaging setup to study the effect of increasing the number of collected events. The image shown in Figure 4D represents a total imaging time of 24 hours (the sum of four 6-hour acquisitions) and contains ~20,000 events.

Timing methods:

The difference in arrival time of the two annihilation photons was determined on an event-by-event basis by the trigger time difference ($t_A - t_B$) between the two digitized signals. In the point source experiments, the trigger time difference of each signal pair was calculated using two different timing methods. The first method was the constant fraction discrimination (CFD) method, which is a conventional method to compute the trigger time of each signal and was implemented in software with parameters of a 0.3 fraction and a 110-ps delay.

For the second method, an alternative to the CFD method was developed to estimate the time-of-flight differences using a convolutional neural network (CNN) (*24*). In this method, the CNN was trained to estimate time-of-flight for each coincidence event directly from the pair of digitized waveforms (*19*). Here, a 9-layer CNN (Fig. S3, full CNN architecture provided in Table S1) was trained for TOF estimation using MATLAB with approximately 1 million coincidence events and stochastic gradient descent with momentum. The training dataset consisted of events measured from a centrally located Na-22 point source, shifted by random timing delays. Layer weights were initialized using He initialization (*25*). Complete training parameters for the CNN are provided in Table S2. Each trigger time difference ($t_A - t_B$) respectively computed by the CFD and CNN methods was histogrammed to form a timing spectrum. The FWHM of each timing spectrum was obtained by linear interpolation of the data points and represented as the coincidence timing resolution.

After comparing the results of the point source experiments using CFD and CNN methods (Fig. 2B), we selected the CNN method to apply to the acquired image phantom data. For timing estimation using the trained CNN with the acquired image phantom data, 4D arrays containing all waveform pairs for each detector line-of-response were input to the CNN. The predicted TOF values output from the CNN were used to generate dPEI images. None of the data used to train the CNN was subsequently reused to determine the timing resolution in Fig. 2B,C or for the imaging data in Fig. 4.

Image generation and post processing

The dPEI setup consists of two single-channel CRI-MCP-PMT detectors, therefore the cross-sectional 2-D image was obtained by building up the dPEI image one row at a time - acquiring a timing spectrum at each *x*-position (Fig. 3) with the collimation and step size described in the "Datasets" section. At each *x*-position, the time-of-flight difference ($t_A-t_B$) of each coincidence event was determined by the CNN method and directly converted to the spatial image domain (without a reconstruction algorithm) using the following equation:

$$y = \frac{c \times (t_A - t_B)}{2} \quad (S1)$$

where $c$ is the speed of light, and $y$ is defined relative to the mid-point of the line-of-response (Fig. 3B). The computed *y*-positions were histogrammed to form one image row (Fig. 3B,C) with a bin size of 1-mm for the dPEI images of the image quality phantom and the spatial resolution test phantom or a bin size of 4-mm for the brain phantom image. The concatenation of all single row images generates the 2-D dPEI images as shown in Fig. 3D.

As described above, dPEI images were generated directly from the measured data without any reconstruction. Subsequently, the image was corrected for attenuation of the 511 keV photons. The probability that both 511-keV photons will reach the detector, $P_{det}$, is given by:

$$P_{det} = \sum_i e^{-\mu_i L_i} \quad (S2)$$

where $L_i$ is the length traversed through material $i$, and $\mu_i$ is the linear attenuation coefficient of material $i$ at 511 keV, which was taken as 0.0969 cm$^{-1}$ for water and 0.1120 cm$^{-1}$ for acrylic. The correction is carried out through manual segmentation of each material type in the non-attenuated corrected dPEI image (*26*). It was assumed that the NEMA NU-4 image quality phantom (except the air-filled void) and the brain phantom were uniformly filled with water, and that the spatial resolution test phantom was a uniform acrylic disk. The attenuation corrected image of the Hoffman brain phantom is shown in Fig. S4B. No correction for accidental coincidences or scattered coincidence events were made, as the contribution of these events in this experimental geometry was determined to be negligible.

After attenuation correction, Gaussian smoothing and image up-scaling were performed to average and interpolate the local pixel values for improved visual representation, reducing the effect of pixel-to-pixel variation. A sigma value of 1.4 pixels (= 1.4 mm) was used for the image quality phantom and the spatial resolution test phantom, and a sigma value of 0.8 pixel (= 3.2 mm) was used for the brain phantom. The same 4-fold up-sampling with bicubic interpolation was performed for all images. Figs. S4C,D show the images after the Gaussian smoothing and rescaling, respectively.

**Supplementary Text**

Accuracy and precision of timing difference with source location
Fig. S5 shows the location of a radioactive point source as determined by the timing difference $t_A - t_B$ (Eq. S1) versus the known location of the source, across a distance of 10 cm. The source location is accurately determined over the entire range.

Effect of number of detected events
Fig. S6 shows images of the 2-D Hoffman brain phantom using differing number of events, ranging from 10,000 to 40,000. This image demonstrates the relatively modest number of detected events needed to form an image of a slice representing the human brain, with little improvement above 20,000 events.

Pathways to increase detection sensitivity of dPEI and reduce activity and acquisition time
This initial demonstration of the principles of dPEI and its implementation should be viewed from the same standpoint as the earliest computed tomography experiments by Hounsfield that yielded the first cross-sectional images (*27*). These experiments involved a radiation source (either a radioactive source or an x-ray tube), and radiation detector that were translated, and also rotated, to produce the necessary projection data for the CT reconstruction. Acquisition times were initially as long as 9 days. Now, of course, CT scans of a large volume of tissue can be accomplished at low radiation doses in well under a second. One can readily imagine a similar development path for dPEI, by increasing the efficiency of the detectors, and also by developing detectors that can be tiled together into linear or 2-D arrays to increase geometric coverage thus allowing large numbers of photon paths through the object to be measured simultaneously.

In the main text, we state that such changes alone could easily increase the detection sensitivity by a factor of $>10^3$, reducing acquisition times or radiation doses accordingly. Here we provide the basis for that assertion. First, by replacing the relatively low efficiency lead glass Cerenkov radiator used in the present work, with, for example, a 4.2 mm thick bismuth germanate (which is both an efficient Cherenkov radiator and a scintillator) entrance window, the interaction cross-sections predict a 6.7-fold detection sensitivity gain over the current lead glass integrated CRI-MCP-PMTs at an energy of 511 keV. This does not account for the fact that BGO, with its higher refractive index, produces more Cherenkov light than the lead glass, and would thus likely lead to a further efficiency gain by producing more events that lie above the signal threshold that is set to remove electronic noise. In the case of the brain phantom scan, the current dPEI system consists of single channel detectors with a sensitive area of 8 × 10 mm$^2$. By developing multi-channel detectors with a sensitive area of 23 × 50 mm$^2$, which are currently undergoing a feasibility design study, the detection sensitivity gain from increasing the geometrical coverage will be 206-fold. Combining both developments will improve the detection sensitivity to 1,380-fold. Therefore, to acquire the same counts to generate the brain phantom dPEI image shown in Fig. 4C, the scan time will be dramatically reduced from 1,440 minutes (24 hours) to only ~1 minute. Clearly the detection area and detection efficiency could be increased even further by tiling multiple such detectors together to form larger area arrays that cover whole organs or ultimately the whole-body. While much technological development is needed to reach these milestones, there are no physical principles that prevent these very large gains in detection efficiency which can ultimately translate into rapid and low-dose dPEI in three dimensions.

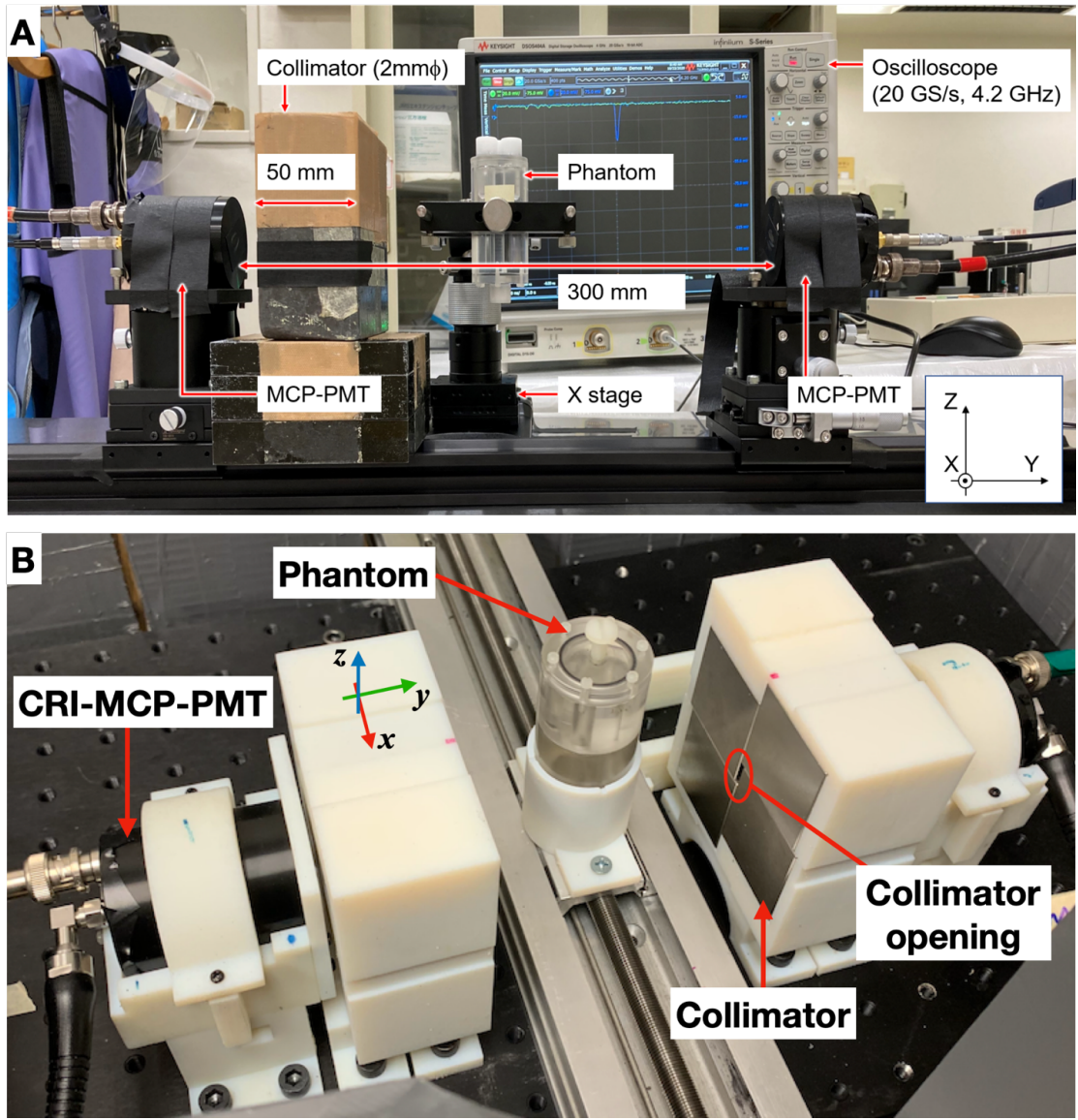

**Fig. S1. Experimental dPEI scan setup used at the two independent sites**. (A) at Hamamatsu Photonics K.K. and (B) at the University of California Davis.

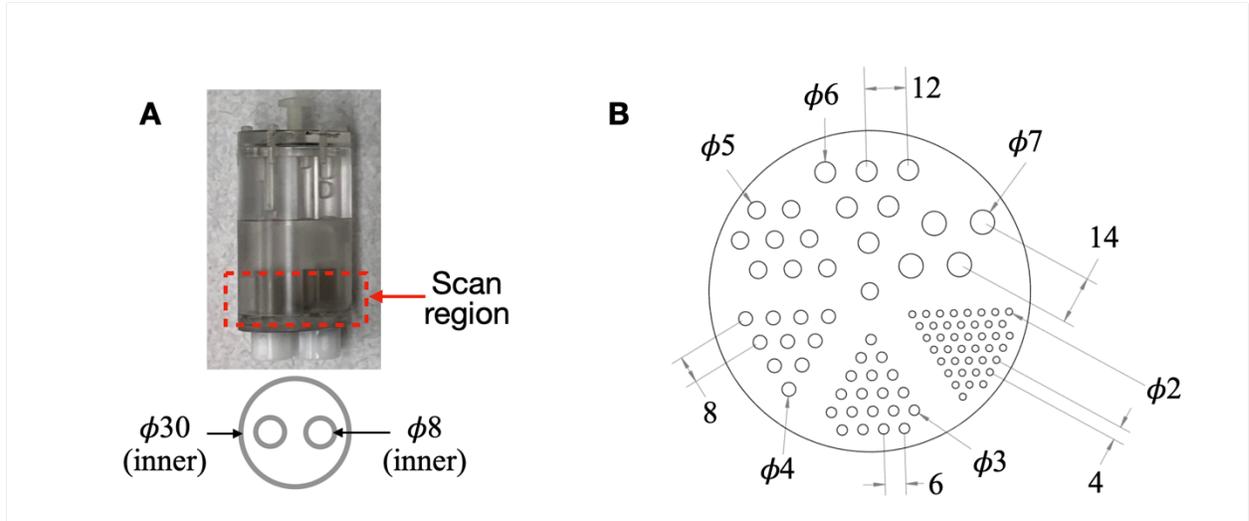

**Fig. S2. Geometry of test objects used for imaging.** Schematics of (A) NEMA NU-4 preclinical image quality phantom and (B) Derenzo spatial resolution phantom used to characterize imaging dPEI performance.

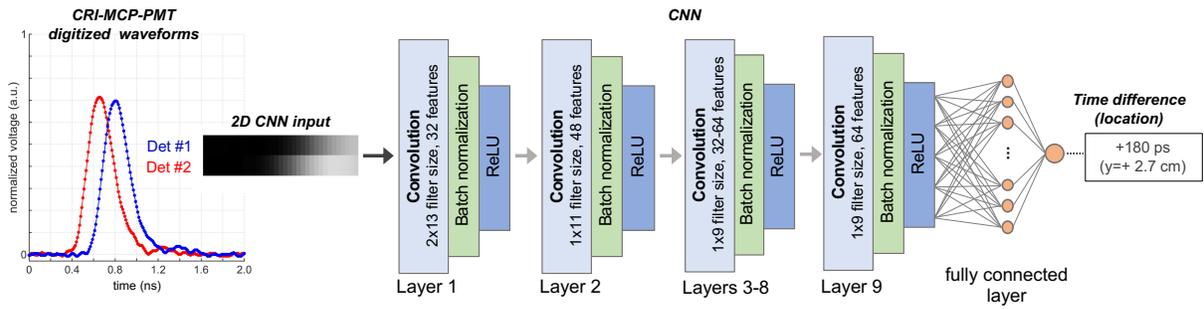

**Fig. S3. CNN architecture.** The CNN used for estimating detection time differences between coincidence annihilation photons. The input to the CNN is the pair of digitized waveforms measured for each coincidence event. Detailed CNN parameters are provided in Tables S1 and S2.

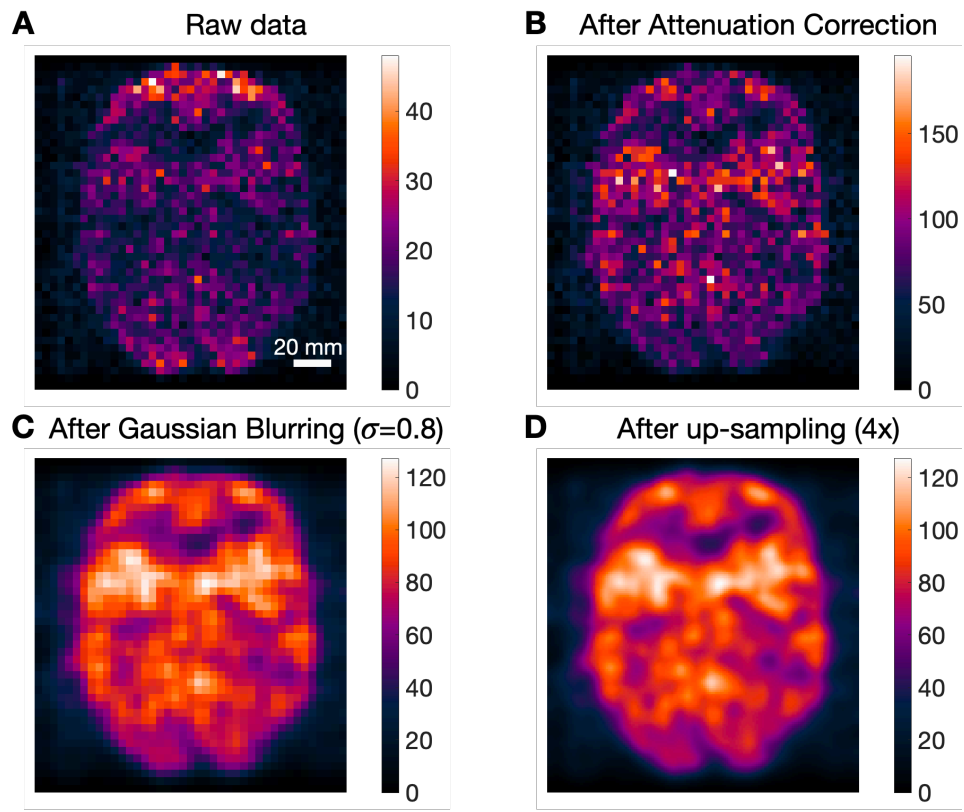

**Fig. S4. dPEI image generation and post-processing steps.** Sequential steps shown as applied to the 2-D Hoffman brain phantom: (A) raw data with no additional processing, (B) after applying attenuation correction, (C) Gaussian blurring with a sigma of 0.8, and (D) 4-fold up-scaling of the image.

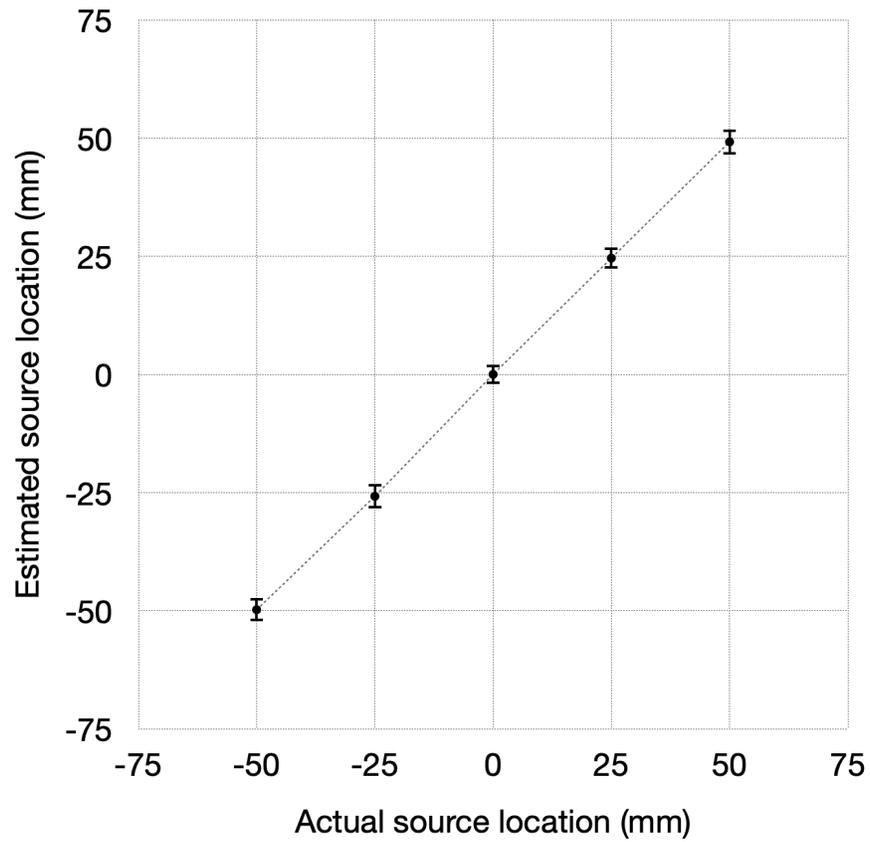

**Fig. S5. Accuracy of source localization based on measured timing difference.** Plot of estimated source location based on timing versus actual location using the data in Figure 2C. Error bar represents ± (FWHM of the distribution at each location ÷ 2).

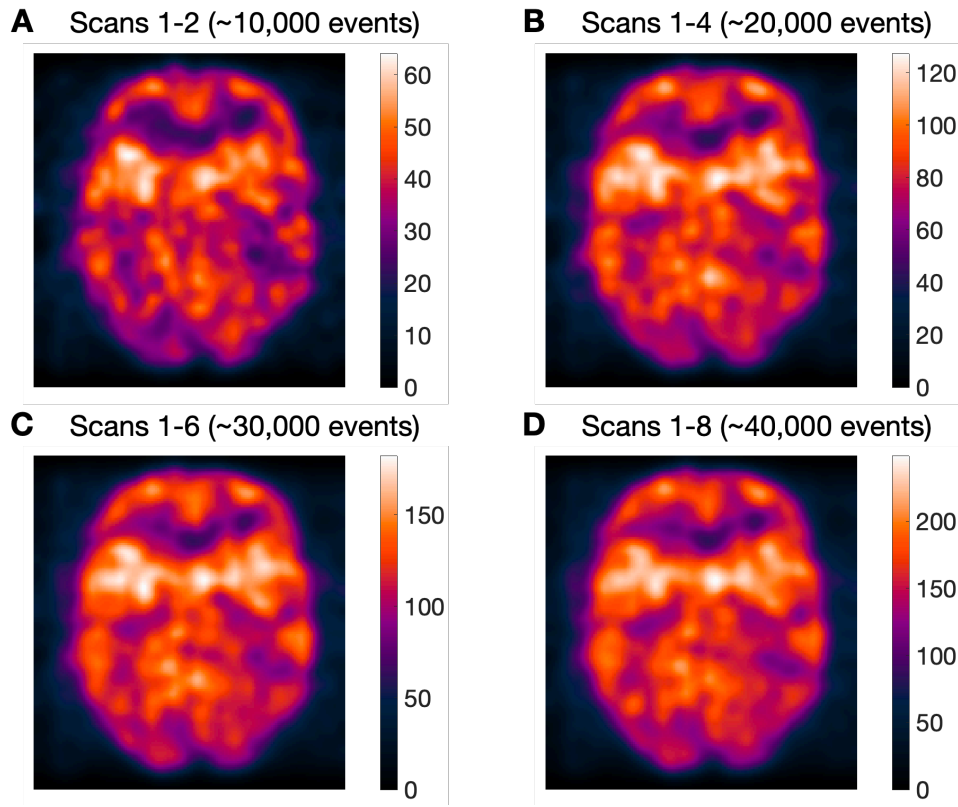

**Fig. S6. Effect of number of detected events on dPET images.** dPEI images of the 2-D Hoffman brain phantom generated using a different number of events: (A) ~10,000 events, (B) ~20,000 events, (C) ~30,000 events, and (D) ~40,000 events. Each acquisition was performed over 44 different *x*-positions (4-mm intervals) and each scan took a total of 6 hours and used ~850 MBq (~23 mCi) of $^{18}$F-FDG activity. All images were post-processed (analytical attenuation correction, Gaussian smoothing ($\sigma$=0.8), and 4-fold up-sampling) as shown in Fig. S4.

| Image input layer | 2 × 400 | |
|---|---|---|
| 2D convolution | 2 × 13 filter size | 16 features |
| Batch normalization | | |
| ReLU | | |
| 2D convolution | 1 × 11 filter size | 32 features |
| Batch normalization | | |
| ReLU | | |
| 2D convolution | 1 × 9 filter size | 32 features |
| Batch normalization | | |
| ReLU | | |
| 2D convolution | 1 × 9 filter size | 32 features |
| Batch normalization | | |
| ReLU | | |
| 2D convolution | 1 × 9 filter size | 32 features |
| Batch normalization | | |
| ReLU | | |
| 2D convolution | 1 × 9 filter size | 48 features |
| Batch normalization | | |
| ReLU | | |
| 2D convolution | 1 × 9 filter size | 48 features |
| Batch normalization | | |
| ReLU | | |
| 2D convolution | 1 × 9 filter size | 64 features |
| Batch normalization | | |
| ReLU | | |
| 2D convolution | 1 × 9 filter size | 64 features |
| Batch normalization | | |
| ReLU | | |
| Fully connected layer | 16 output neurons | |
| ReLU | | |
| Fully connected layer | 1 output neuron | |
| Regression prediction | | |

**Table S1. CNN layers**. (ReLU: rectified linear activation function).

| | |
|---|---|
| **Learning rate** | 0.0001 |
| **Number of epochs** | 3 |
| **Mini-batch size** | 32 |
| **Learning rate drop factor (epoch)** | 0.3 |
| **L2 regularization** | 0.00001 |
| **Cost function** | Mean squared error |

**Table S2.** Parameters used in training the CNN for timing estimation.